\documentclass[conference]{IEEEtran}
\IEEEoverridecommandlockouts
\usepackage{cite}
\usepackage{amsmath,amssymb,amsfonts}
\usepackage{algorithmic}
\usepackage{graphicx}
\usepackage{textcomp}
\usepackage{xcolor}
\usepackage{listings}
\usepackage{multirow}

\usepackage{graphicx}
\usepackage{subcaption}
\usepackage{hyperref}

\usepackage[most]{tcolorbox}
\newtcolorbox{promptsimple}{
	colback=blue!5,            
	colframe=blue!40!white,
	arc=4pt,                   
	boxrule=0.5pt,             
	left=6pt,right=6pt,top=6pt,bottom=6pt,
	boxsep=2pt,
	enhanced,
	breakable
}

\def\BibTeX{{\rm B\kern-.05em{\sc i\kern-.025em b}\kern-.08em
    T\kern-.1667em\lower.7ex\hbox{E}\kern-.125emX}}
\begin{document}

\makeatletter
\newcommand{\linebreakand}{%
  \end{@IEEEauthorhalign}
  \hfill\mbox{}\par
  \mbox{}\hfill\begin{@IEEEauthorhalign}
}

\makeatother

\title{LLM-Driven Heuristic Frame-Level Quantization Parameter Adaptation for VVenC
}

\author{
\IEEEauthorblockN{
Liqiang He$^{1\ast}$,
Yingwen Zhang$^{1\ast}$,
Riyu Lu$^{1,2}$,
Meng Wang$^{3}$,
Shiqi Wang$^{1\ddagger}$
}
\IEEEauthorblockA{
$^{1}$Department of Computer Science, City University of Hong Kong\\
$^{2}$Department of Computer Science and Technology, Harbin Institute of Technology\\
$^{3}$School of Data Science, Lingnan University \\
}

\thanks{$\ast$ equal contributions. $\ddagger$ corresponding: shiqwang@cityu.edu.hk}

}

\maketitle

\begin{abstract}
Optimal frame-level quantization parameter (QP) allocation remains a persistent challenge in modern video encoders. The fixed-QP scheme widely adopted in practical systems is inherently content-agnostic, while classical Lagrangian rate-distortion optimization (RDO) methods often suffer from inaccurate multiplier settings. In this paper, we explore the use of large language models (LLMs) to automatically design RDO heuristics for frame-level QP adaptation. We construct a closed-loop evolutionary framework in which the LLM iteratively proposes RDO heuristics as algorithmic ideas with executable code, and these candidates are evaluated directly through encoding with the Fraunhofer Versatile Video Encoder (VVenC), where each heuristic acts as a scoring function that compares different QP choices based on the encoding statistics of past frames and current candidates. Experimental results across multiple test sets show that the evolved heuristic achieves promising rate-distortion improvements over both the fixed-QP scheme and the Lagrangian baseline. Further analysis reveals that the LLM can autonomously discover an adaptive heuristic that penalizes QP fluctuations via entropy-based terms, providing new insights into the design of RDO algorithms.
\end{abstract}

\section{Introduction}
\label{sec:intro}
Among the various encoder implementations of Versatile Video Coding (VVC)~\cite{bross2021overview}, the Fraunhofer Versatile Video Encoder (VVenC)~\cite{VVENC} stands out as a highly optimized, open-source encoder that effectively balances compression performance with practical usability. 
However, achieving optimal rate-distortion (RD) performance for VVenC remains a complex challenge. This is largely due to the highly complex parameter space of modern encoders, which involves numerous interdependent coding parameters. Among these parameters, the frame-level Quantization Parameter (QP) is one of the most critical parameters, as it directly determines bit allocation across frames, thereby affecting the trade-off among overall bitrate, distortion, and subjective visual quality.

In the hierarchical random access (RA) configuration, a fixed set of QP offsets is assigned to frames across different temporal layers~\cite{schwarz2005hierarchical}, reflecting a well-established bit allocation principle: lower QPs for frequently referenced frames and higher QPs for those less critical to subsequent prediction. VVenC adopts the same fixed scheme for frame-level bit allocation. While effective in many scenarios, this approach implicitly assumes that frames at the same temporal layer share similar statistics. In practice, texture complexity, motion magnitude, and reference propagation can vary significantly across sequences. The fixed QP allocation scheme in VVenC is therefore suboptimal, leaving clear room for content-adaptive improvement.

In this paper, we attempt to address this issue by leveraging a classic technique in video coding: rate-distortion optimization (RDO)~\cite{sullivan1998rate}. The core philosophy of RDO is to define a proper heuristic or measure~\cite{zhang2025theoretical}, typically the Lagrangian cost \(J = \lambda R + D\), that quantifies the ``goodness'' of each coding decision, thereby enabling direct comparison through the heuristic score. For instance, a smaller score of $J$ usually indicates a better decision. However, applying such a Lagrangian heuristic to frame-level QP adaptation remains problematic~\cite{Riyu_1,Riyu_2}, primarily because an accurate \(\lambda\) is difficult to obtain. In principle, deriving a proper \(\lambda\) demands precise RD modeling~\cite{li2014lambda}, temporal distortion propagation modeling~\cite{li2016lambda}, and highly content-adaptive parameter updating, all of which are non-trivial in practice.

\begin{figure*}
    \centering
    \includegraphics[width=0.9\linewidth]{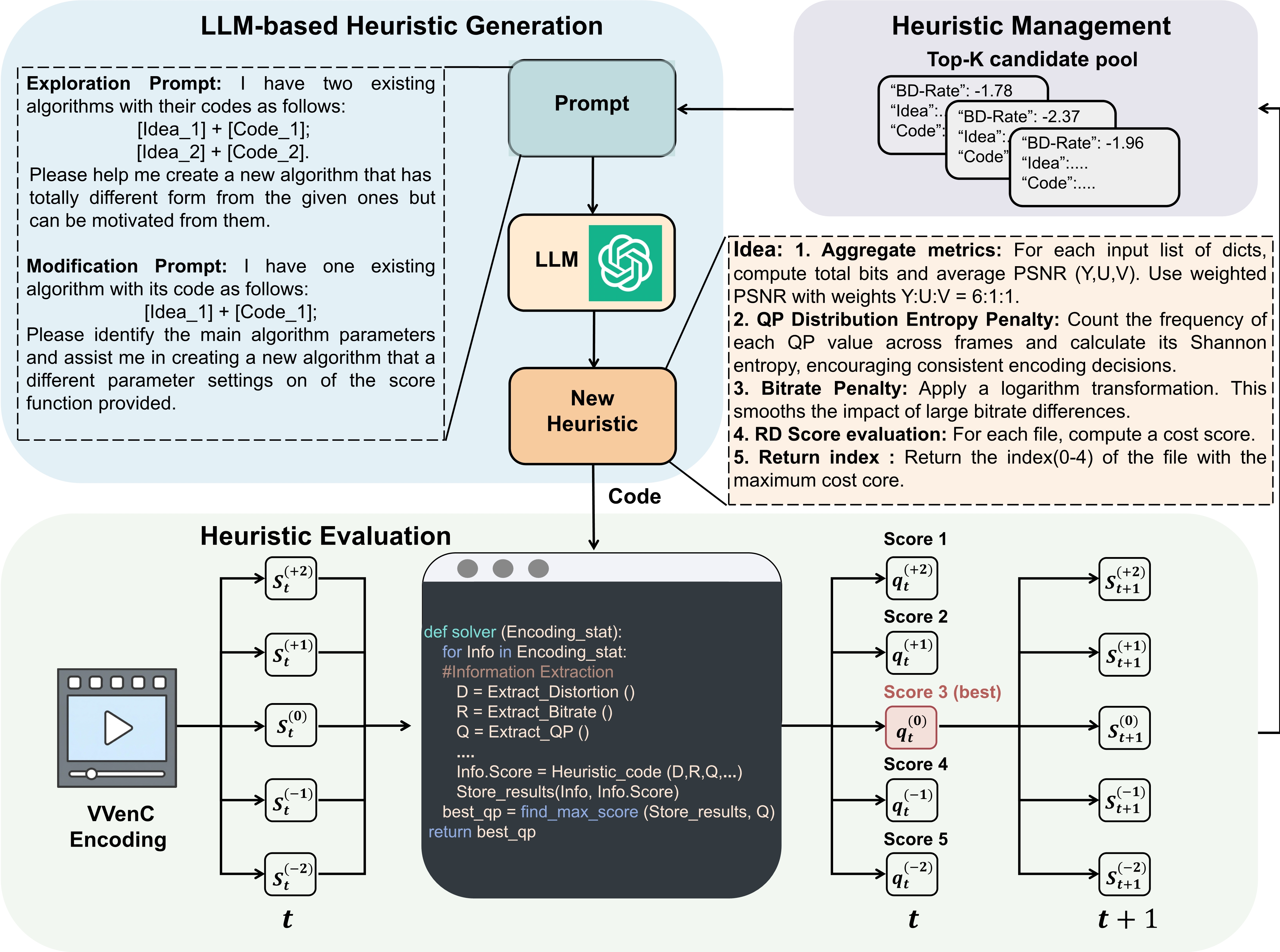}
    \caption{Proposed LLM-driven RDO heuristic design framework for frame-level QP adaptation.}
    \label{fig:framework}
\end{figure*}

Recently, Large Language Models (LLMs) have rapidly advanced beyond natural language processing and expanded into domains such as programming assistance~\cite{LLM_programming}, biomedical engineering~\cite{LLM_Medicine}, and chemical discovery~\cite{LLM_chemistry}. They have demonstrated remarkable capability to automatically generate novel heuristics through question-and-answer interactions~\cite{liu2024evolution}. In this paper, we explore the potential of LLM-driven RDO heuristic design for frame-level QP adaptation. Specifically, we build a closed‑loop optimization framework based on LLMs in which an RDO heuristic, comprising both algorithmic ideas and executable code, is iteratively generated and evaluated. At each iteration, the LLM is prompted to generate new RDO heuristics via question-and-answer, conditioned on historical heuristics. The generated heuristic is then deployed in frame-level QP RDO. The resulting compression performance, measured by the Bjøntegaard Delta Rate (BD-rate)~\cite{bjontegaard2001calculation}, is used by the LLM as an objective to judge the heuristic’s quality in subsequent iterations. Experimental results show that such an evolved heuristic outperforms both the VVenC fixed‑QP scheme and the Lagrangian heuristic. Interestingly, as an example of such independently discovered heuristics, we provide a detailed analysis of an insightful entropy-constrained RDO heuristic that offers meaningful guidance for future RDO heuristic design. To the best of our knowledge, this work presents the first LLM‑driven heuristic frame‑level QP adaptation method in video coding.

\section{LLM-Driven RDO Heuristic Design}

\subsection{Preliminaries: RDO Heuristic}\label{sec:preliminaries}
A typical frame-level RDO algorithm is summarized as follows, which has been successfully employed for frame-level coding scale adaptation~\cite{Riyu_2}. In this work, we extend it to the frame-level QP adaptation scenario. Consider a video sequence of length \(T\). Let the QP vector 
\begin{equation}\label{eq:candidate}
\mathbf{Q} = [q_0, q_1, \dots, q_{T-1}]
\end{equation}
denote the frame-level QP decisions in coding order, where \(q_t\) is the QP assigned to the \(t\)-th frame. At time step \(t\), a base QP value is first determined, and a candidate set centered at this value is generated as
\begin{equation}
\mathcal{Q}_t = \left\{ q_t^{(-2)},\; q_t^{(-1)},\; q_t^{(0)},\; q_t^{(+1)},\; q_t^{(+2)} \right\},
\end{equation}
where \(q_t^{(0)}\) is the base QP and the other candidates deviate by, e.g., \(\pm1\) and \(\pm2\) (\(|\mathcal{Q}_t| = 5\)). Subsequently, the current frame is encoded with each candidate \(q_t^{(k)} \in \mathcal{Q}_t\), yielding a bitrate \(R(q_t^{(k)})\) and distortion \(D(q_t^{(k)})\). The Lagrangian cost is then computed as 
\begin{equation}\label{eq:lagrangian-heuristic}
J(q_t^{(k)}) = D(q_t^{(k)}) + \lambda_t \cdot R(q_t^{(k)}),
\end{equation}
where \(\lambda_t\) is a Lagrange multiplier pre-determined by the VVenC codec. Based on such a heuristic score, the QP for the current frame is selected as 
\begin{equation}\label{eq:rdo}
q_t^* = \arg\min_{q_t^{(k)} \in \mathcal{Q}_t} J(q_t^{(k)}),
\end{equation}
where a smaller score indicates a better decision. This process is performed sequentially following the RA coding order. The resulting QP vector 
\begin{equation}
\mathbf{Q}^* = [q_0^*, q_1^*, \dots, q_{T-1}^*]
\end{equation}
constitutes the final frame-level QP allocation. This classic RDO heuristic provides a simple baseline, but its performance depends heavily on the predefined \(\lambda_t\), which should ideally be context-adaptive but is not in practice, motivating our exploration of more flexible, LLM-generated heuristics.

\subsection{Proposed Framework}

To this end, we propose an LLM-driven closed-loop evolutionary framework that automates the optimization of RDO heuristics. As illustrated in Fig.~\ref{fig:framework}, the framework operates as an iterative evolutionary process comprising three core modules: LLM-based heuristic generation, heuristic evaluation, and heuristic management. The LLM generates RDO heuristics, which are then evaluated via VVenC encoding, and the resulting BD-rate guides the update of the candidate pool for the next generation. This process repeats until the maximum number of iterations is reached.

\textbf{LLM-Based Heuristic Generation.} As the core engine of the framework, this module uses LLMs to generate new heuristics guided by historical evaluation results. At each iteration, task-specific prompts are constructed and fed to the LLM to produce novel heuristics. Two evolutionary prompt strategies~\cite{liu2024evolution} are adopted: exploration and modification. Exploration encourages large-scale structural variations, enabling the current heuristic to escape local optima and discover qualitatively different design patterns. Modification, by contrast, performs localized parameter-level fine-tuning of promising candidates within a narrower search radius. Each prompt is conditioned on historical heuristics through a structured representation: an algorithmic idea that captures the core design rationale, paired with its executable code that can be directly embedded into the RDO pipeline. The LLM's output is also constrained to follow the same idea-code template. Within each iteration, multiple rounds of exploration and modification are issued, producing a batch of candidate heuristics that are subsequently evaluated through VVenC encoding runs. These two prompt strategies, exploration and modification, are illustrated in Fig.~\ref{fig:framework}, along with an example of a generated idea, while Listing~\ref{code:solver} shows the corresponding code. For the initial population, randomly generated seeds produced by the LLM are used.

\textbf{Heuristic Evaluation.} This module quantifies the ``goodness'' of an LLM-generated heuristic by applying it to frame-level QP RDO and measuring the resulting BD-rate against the fixed-QP scheme. Unlike the Lagrangian heuristic defined in Eqn.~(\ref{eq:lagrangian-heuristic}), which only considers the bitrate and distortion of the current frame (step \(t\)), our LLM-generated heuristic takes as input the encoding statistics of all frames from step \(0\) to step \(t\), thereby enabling the LLM to freely explore and autonomously discover effective decision rules beyond traditional Lagrangian heuristics. Specifically, let
\begin{equation}
\mathcal{S}_t = \left\{ S_t^{(-2)}, S_t^{(-1)}, S_t^{(0)}, S_t^{(+1)}, S_t^{(+2)} \right\},
\end{equation}
denote the encoding statistics associated with candidate QPs in $\mathcal{Q}_t$ at step \(t\), where each \(S_t^{(k)}\) includes both the historical statistics of previous frames (e.g., bitrate, distortion, and QP decisions) and the encoding results of the current frame when encoded with the \(q_t^{(k)}\). Then, the LLM-generated heuristic is implemented as an executable function. It takes \(\mathcal{S}_t\) as input and returns the index of the candidate that its internal logic considers optimal. Formally, the decision rule becomes:
\begin{equation}\label{eq:heuristic-rdo}
q_t^* = \arg\max_{q_t^{(k)} \in \mathcal{Q}_t} {H}(\mathcal{S}_t),
\end{equation}
where \({H}\) denotes the heuristic function generated by the LLM. This decision-making process is repeated sequentially for all frames in the sequence until the last frame is reached. After the entire sequence is encoded, the resulting BD-rate score serves as the fitness signal, which is fed back to the candidate pool to guide subsequent evolutionary refinement.

Implementing the heuristic RDO defined in Eqn.~(\ref{eq:heuristic-rdo}) presents a challenge: it typically requires substantial modifications to the encoder's core RDO functionality. To address this issue without deep engineering, we adopt the multi-pass encoding strategy proposed in~\cite{Riyu_1}. Specifically, for each QP candidate \(q_t^{(k)} \in \mathcal{Q}_t\), we encode the current frame using the cached history of QP decisions from frames \(0\) to \(t-1\). After completing all \(|\mathcal{Q}_t|\) encoding runs, we collect their statistics into $\mathcal{S}_t$. This allows the entire heuristic RDO score calculation and candidate comparison to be performed outside the encoder, relying only on standard encoding logs, enabling a clean separation between heuristic definition and codec implementation. Further details are available in~\cite{Riyu_1}.

\textbf{Heuristic Management.} This module maintains a fixed-size population of heuristics and serves as the memory and selection mechanism of the closed-loop framework. Its primary responsibilities are twofold: retaining high-performing heuristics discovered during evolution and providing algorithmic parents for subsequent prompt generation. Formally, the population $\mathcal{P}$ is defined as an ordered list of fixed capacity $K$:
\begin{equation}
\mathcal{P} = \{(C_1, I_1, b_1), (C_2, I_2, b_2), \ldots, (C_K, I_K, b_K)\},
\end{equation}
where $C_i$, $I_i$, and $b_i$ denote the executable heuristic code, its corresponding idea, and the evaluated BD-rate, respectively. The population is maintained in descending order of BD-rate, with $(C_1, I_1, b_1)$ being the best individual discovered so far.

At each iteration, newly generated heuristics are evaluated and assigned a score \(b_{\text{new}}\). If \(b_{\text{new}}\) surpasses \(b_K\), the new heuristic is inserted at the appropriate rank, and the worst-performing heuristic is discarded. This ensures consistent improvement across iterations. The high-performing heuristics in \(\mathcal{P}\) are then sampled as parents based on their BD-rate values (i.e., better BD-rate leads to higher sampling probability), with their code, idea, and metrics injected into the prompt templates (Fig.~\ref{fig:framework}) to guide the LLM in building upon proven algorithmic patterns.

\section{Experimental Results}
\subsection{Experimental Settings}
We use DeepSeek~\cite{liu2025deepseek}, a frontier open-source model, as the LLM engine to validate our framework. Our experimental encoder is VVenC-1.7.0~\cite{VVENC} with the faster preset under the default RA configuration. The greater practicality of VVenC makes it suitable for large-scale evaluation of LLM-generated heuristics. The intra period is 32 and the encoding frame length is 65, applied to all experiments. YUV-PSNR is computed with a Y:U:V ratio of 6:1:1. Coding efficiency is measured by BD-rate, where negative values indicate bitrate savings relative to the anchor. Our target resolution is 1080p. 
\begin{table}[htbp]
\caption{BD-rate (YUV-PSNR) comparison under RA configuration. The anchor is the fixed-QP scheme.} \label{tab:BD_Rate}
\setlength{\tabcolsep}{8pt}
\centering
\renewcommand{\arraystretch}{1.4}
\begin{tabular}{|cc|c|}
\hline
\multicolumn{2}{|c|}{\textbf{Sequences}}                                                                                            & \textbf{Proposed Method} \\ \hline
\multicolumn{1}{|c|}{\multirow{6}{*}{\textbf{\begin{tabular}[c]{@{}c@{}}CTC\\  Class B\end{tabular}}}} & BasketballDrive & -2.39\%                  \\ \cline{2-3} 
\multicolumn{1}{|c|}{}                                                                                   & BQTerrace       & -0.05\%                  \\ \cline{2-3} 
\multicolumn{1}{|c|}{}                                                                                   & Cactus          & -2.31\%                  \\ \cline{2-3} 
\multicolumn{1}{|c|}{}                                                                                   & MarketPlace     & -1.85\%                  \\ \cline{2-3} 
\multicolumn{1}{|c|}{}                                                                                   & RitualDance     & -2.07\%                  \\ \cline{2-3} 
\multicolumn{1}{|c|}{}                                                                                   & \textbf{Average}         & \textbf{-1.73\%}                  \\ \hline
\multicolumn{2}{|c|}{\textbf{UVG}}                                                                                                  & \textbf{-1.00\%}                  \\ \hline
\multicolumn{2}{|c|}{\textbf{MCL-JCV}}                                                                                              & \textbf{-1.06\%}                  \\ \hline
\end{tabular}
\end{table}For the LLM-driven heuristic evolution stage, we encode five sequences\footnote{\scriptsize Wood, TrafficAndBuilding, TrafficFlow, TallBuildings, ResidentialBuilding.} from the SJTU 4K dataset~\cite{song2013sjtu} (downsampled to 1080p) to obtain the BD-rate against the fixed-QP scheme of each candidate heuristic. We perform 10 iterations, each generating 4 heuristic candidates. In each iteration, we apply both exploration and modification prompts twice, resulting in a total of 4 candidates. The candidate pool size is fixed to \(K=4\), and the number of parent heuristics is 2. The best-scoring heuristic from the final generation is selected and fixed for all testing. For the testing dataset, we report coding performance on CTC Class B, UVG~\cite{mercat2020uvg}, and MCL-JCV~\cite{wang2016mcl}. QP values are set to 22, 27, 32, and 37. For each target QP value, at every time step the base QP is set to that fixed QP, and the candidate set \(\mathcal{Q}_t\) is generated by adding offsets \(\pm1, \pm2\) to the base QP, i.e., \(\{q_t^{(-2)}, q_t^{(-1)}, q_t^{(0)}, q_t^{(+1)}, q_t^{(+2)}\}\). In addition to bitrate \(R_i\), distortion \(D_i\), and the QP value itself, we extract other information from the VVenC encoding logs for each frame, such as frame type, reference layer, and filtering strength. These features are aggregated into the historical statistics \(S_t^{(k)}\) (see Sec.~\ref{sec:preliminaries}), allowing the LLM to autonomously decide whether and how to exploit them. We optimize a separate heuristic for each QP value. During the evolution stage, for a given QP, the BD-rate is evaluated by encoding only that QP with the heuristic, while the other three QPs use the fixed‑QP anchor.

\begin{figure*}[!t]
    \centering
    \begin{subfigure}[b]{0.45\textwidth}
        \centering
        \includegraphics[width=\linewidth]{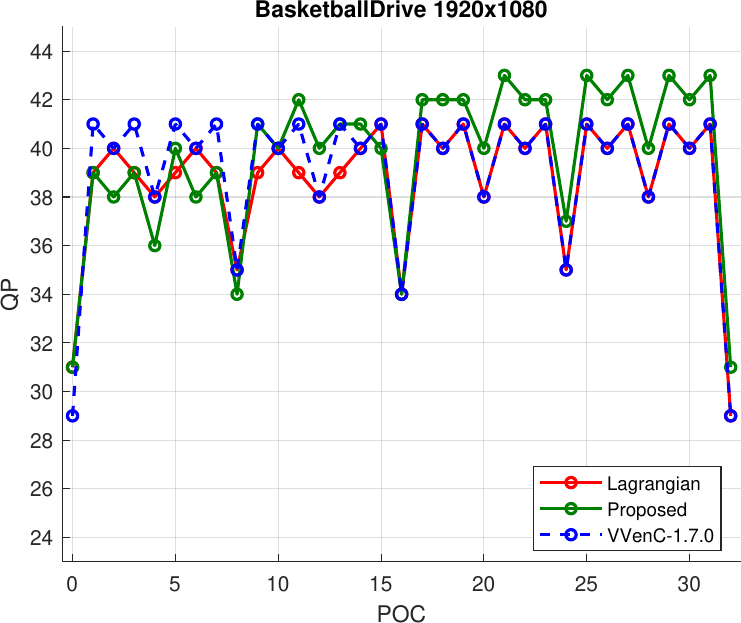}
        \caption{}
        \label{fig:sub1}
    \end{subfigure}
    \hfill
    \begin{subfigure}[b]{0.45\textwidth}
        \centering
        \includegraphics[width=\linewidth]{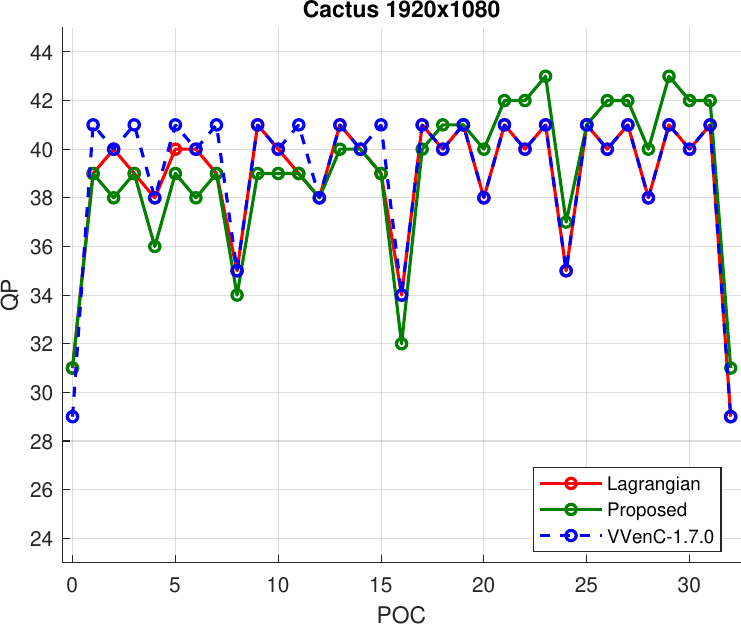}
        \caption{}
        \label{fig:sub2}
    \end{subfigure}
    \caption{QP allocation within one GOP for (a) \textit{BasketballDrive} and (b) \textit{Cactus} at QP=32.}
    \label{fig:both}
\end{figure*}

\begin{lstlisting}[
    language=Python,
    frame=lines,
    basicstyle=\ttfamily\scriptsize,
    numbers=left,
    numberstyle=\tiny\color{gray},
    stepnumber=1,
    numbersep=5pt,
    tabsize=4,
    breaklines=true,
    breakatwhitespace=true,
    caption={The generated heuristic function of QP=32.},
    label={code:solver},
    float=htbp,
    captionpos=b
]
def solver(stat_0: List[Dict[str, Any]], stat_1: List[Dict[str, Any]], stat_2: List[Dict[str, Any]], stat_3: List[Dict[str, Any]], stat_4: List[Dict[str, Any]]) -> int:
    
    all_stats = [stat_0, stat_1, stat_2, stat_3, stat_4]
    scores = []
    
    for stat in all_stats:
        ## calculate average bits and PSNR
        total_bits = sum(frame['bits'] for frame in stat)
        qp_counts = {}
        sum_y, sum_u, sum_v = 0.0, 0.0, 0.0
        count = 0
        
        for frame in stat:
            qp = frame['QP']
            qp_counts[qp] = qp_counts.get(qp, 0) + 1
            sum_y += frame['Y']
            sum_u += frame['U']
            sum_v += frame['V']
            count += 1
            
        avg_y = sum_y / count
        avg_u = sum_u / count
        avg_v = sum_v / count
        weighted_psnr = (6 * avg_y + avg_u + avg_v) / 8
        log_bitrate = math.log(total_bits + 1)
        
        ## calculate QP entropy penalty
        qp_entropy = 0.0
        for qp_count in qp_counts.values():
            prob = qp_count / count
            qp_entropy -= prob * math.log(prob + 1e-10)
        
        qp_penalty = 1.0 / (1.0 + qp_entropy)
        
        ## calculate heuristic score
        rd_score = weighted_psnr * qp_penalty / log_bitrate
        scores.append(rd_score)
    
    best_index = max(range(len(scores)), key=lambda i: scores[i])
    return best_index

\end{lstlisting}

\subsection{RD Performance}\label{sec:R-D analyses}
Table~\ref{tab:BD_Rate} presents the BD-rate results of our evolved heuristic compared against the Lagrangian RDO baseline~\cite{Riyu_1,Riyu_2} under the RA configuration. As shown in Table~\ref{tab:BD_Rate}, our heuristics achieve consistent BD-rate savings across all tested datasets. On the CTC Class B sequences, the average BD-rate saving is -1.73\%, with individual gains ranging from -0.05\% (\textit{BQTerrace}) to -2.39\% (\textit{BasketballDrive}). On the UVG and MCL-JCV datasets, the proposed method delivers average savings of -1.00\% and -1.06\%, respectively. Overall, the evolved heuristic achieves an average BD-rate saving of -1.26\% across all test sequences. Following~\cite{Riyu_1,Riyu_2}, we also implemented the Lagrangian heuristic for comparison. While this method proves effective for the frame‑level coding scale RDO, its straightforward extension to QP decision exhibits unstable behavior, resulting in average BD‑rate increases of 2.68\% on CTC Class B, 3.09\% on UVG, and 2.52\% on MCL‑JCV. This performance drop stems from two fundamental limitations of the per-frame Lagrangian heuristic. First, it uses a fixed Lagrangian multiplier \(\lambda\) determined solely by the \(\lambda\)-QP relationship, which fails to adapt to content-varying characteristics across frames. Second, although the Lagrangian framework can, in principle, account for reference dependencies through distortion propagation~\cite{li2016lambda}, doing so requires precise modeling of how the current frame's distortion affects future frames that reference it. In practice, achieving such accurate propagation modeling is difficult, so the heuristic largely ignores inter-frame dependencies. This is particularly detrimental under the RA configuration, where hierarchical reference structures amplify the impact of suboptimal QP decisions on subsequent frames. Consequently, the Lagrangian heuristic often makes decisions that are no better than, or even worse than, using a fixed-QP scheme. By contrast, the evolved heuristic holistically exploits per-candidate encoding statistics, including bitrate, distortion, and possible reference structure information, to potentially learn an implicit model of reference propagation, which in turn suggests a degree of local adaptivity.

\subsection{Insights from an Evolved Heuristic}
To further demonstrate the superiority of our evolved heuristics and to provide deeper insights into their underlying mechanisms, we examine a representative heuristic evaluation function automatically generated by the LLM for QP=32, one of the four discovered heuristics. As discussed earlier, our RDO heuristic computation and candidate comparison~\cite{Riyu_1} are performed entirely outside the VVenC encoder. This is implemented as a standalone Python function, shown in Listing~\ref{code:solver}, which takes five sets of encoding statistics (each corresponding to a different QP value $q_t^{(k)}$) and returns the index of the best-performing candidate. We distill the evaluation logic as follows. At step $t$, for each candidate $q_t^{(k)} \in \mathcal{Q}_t$ and its corresponding encoding statistics $S_t^{(k)}$, the heuristic defines a score:
\begin{equation}
H(S_t^{(k)}) = \frac{\overline{\text{PSNR}}_{\!w}}{\bigl(1 + H_{\text{QP}}(k)\bigr) \cdot \log\bigl(\sum_{i=0}^{t} R_i + 1\bigr)},
\end{equation}
where $\overline{\text{PSNR}}_{\!w} = (6\bar{Y} + \bar{U} + \bar{V})/8$ is the weighted average of the temporal mean PSNR values of the Y, U, V components over frames $0$ to $t$, as recorded in $S_t^{(k)}$. $R_i$ denotes the bitrate of frame $i$. The term $H_{\text{QP}}(k)$ is the entropy of the QP distribution observed across frames $0$ to $t$ in $S_t^{(k)}$. The heuristic finally selects the candidate with the highest score:
\begin{equation}
k^* = \arg\max_k \; H(S_t^{(k)}).
\end{equation}

Remarkably, the LLM has discovered two simple yet non‑trivial mechanisms solely from the raw encoding statistics $S_t^{(k)}$, without relying on any complex models. First, the score takes a ratio form: $\overline{\text{PSNR}}_{\!w} / \log(\sum R_i+1)$. Because PSNR itself is a logarithmic measure of distortion (PSNR $\propto -\log(\text{MSE})$), both numerator and denominator live in the logarithmic domain. The ratio therefore directly measures the weighted PSNR gain per unit logarithmic bitrate: the heuristic prefers candidates that deliver more distortion reduction for each logarithmically counted bit. However, there is a subtle structural asymmetry: the numerator averages per‑frame PSNR values (each already logarithmic), whereas the denominator first sums the raw bitrates and then takes a single logarithm. This makes the denominator much more sensitive to additional bits when the cumulative bitrate is small, and much less sensitive when it is large.
Consequently, this yields a dynamic allocation strategy that automatically adjusts its sensitivity to bit consumption based on the current cumulative bitrate, without any manually tuned $\lambda$. Second, the entropy penalty $H_{\text{QP}}(k)$ appears in the denominator as $(1+H_{\text{QP}}(k))$. A low entropy (most frames use the same QP) keeps the penalty near 1, whereas a high entropy (frequent QP switching) reduces the score. This discourages unnecessary QP fluctuations, which is particularly beneficial in hierarchical RA coding because unstable QP decisions on reference frames propagate prediction errors. Crucially, both mechanisms rely solely on historical information stored in $S_t^{(k)}$: the cumulative bitrate $\sum R_i$ and the QP distribution over all past frames. In sharp contrast, the vanilla Lagrangian heuristic $J = D + \lambda R$ only looks at the current frame's instantaneous rate and distortion, ignoring all past decisions.

To intuitively reveal the frame-level QP adaptation behavior of the LLM-driven heuristic, Fig.~\ref{fig:both} compares the per-frame QP decisions of the heuristic given in Listing~\ref{code:solver} with the Lagrangian and the fixed-QP offset scheme of VVenC on the sequences \textit{BasketballDrive} and \textit{Cactus} at QP=32. The default VVenC anchor scheme (blue curve) shows a rigid, content-agnostic allocation determined solely by temporal layer hierarchy. The Lagrangian heuristic (red curve) largely follows this pattern but occasionally selects lower QPs for less critical B‑frames (e.g., POCs 1 and 3), which inadvertently allocates bits away from hierarchical reference frames and disrupts temporal dependency propagation, leading to BD‑rate increase. This misbehavior stems from suboptimal Lagrangian multiplier $\lambda$ settings in VVenC, indicating room for further optimization. In contrast, the evolved heuristic (green curve) generally respects the hierarchical principle (lower QPs for reference frames, higher QPs for non‑reference frames) while exhibiting nontrivial flexibility: occasionally, it assigns a low‑layer frame the same QP as a high‑layer frame, as seen at POC~18 in \textit{BasketballDrive} and POC~10 in \textit{Cactus}. This selective relaxation of the strict hierarchy allows the heuristic to better adapt to local coding conditions without the handcrafted constraints of the Lagrangian approach.

\section{Conclusion}
In this paper, we presented an LLM-driven evolutionary framework that discovers effective RDO heuristics for frame-level QP adaptation in VVenC. The evolved heuristics outperform both the fixed-QP scheme and the Lagrangian baseline, achieving consistent BD-rate savings of around 1--2\% across diverse test sets. Detailed analysis of a representative heuristic further demonstrates that the LLM can identify non-trivial mechanisms without any handcrafted complex model. These results suggest the potential of LLMs as automatic optimizers for video codecs. Future work will extend the framework to broader codec optimization tasks.
 

\end{document}